\documentclass[9pt,twocolumn,twoside]{opticajnl}
\usepackage{soul}

\journal{ol} 

\setboolean{shortarticle}{true}

\title{Visible-telecom broadband optical isolator based on dynamic modulation in thin-film lithium niobate}

\author[1]{Manav Shah}
\author[1]{Ian Briggs}
\author[1]{Pao-Kang Chen}
\author[1]{Songyan Hou}
\author[1,*]{Linran Fan}

\affil[1]{James C. Wyant College of Optical Sciences, The University of Arizona, 1630 E. University Blvd., Tucson, AZ 85721, USA}

\affil[*]{Corresponding author: lfan@optics.arizona.edu}

\begin{abstract}
Optical isolators are an essential component of photonic systems. Current integrated optical isolators have limited bandwidths due to stringent phase-matching conditions, resonant structures, or material absorption. Here, we demonstrate an ultra-broadband integrated optical isolator in thin-film lithium niobate photonics. We use dynamic standing-wave modulation in a tandem configuration to break Lorentz reciprocity and achieve isolation. We measure an isolation ratio of 15 dB and insertion loss below 0.5 dB for a design wavelength of 1550 nm. In addition, we experimentally show that this isolator can simultaneously operate at visible and telecom wavelengths with comparable performance. Isolation bandwidths $\sim$100~nm can be achieved simultaneously at both visible and telecom wavelengths. Our device's large bandwidth, high flexibility, and real-time tunability can enable novel non-reciprocal functionality on integrated photonic platforms.
\end{abstract}

\setboolean{displaycopyright}{true}

\begin{document}

\maketitle
Optical isolators facilitate unidirectional optical transmission and are widely used to suppress optical back-reflection. This functionality is critical in stabilizing laser cavities, protecting high-power optical amplifiers, and preventing multi-path optical interference. Optical isolators are characterized by an asymmetric scattering matrix that breaks Lorentz reciprocity \cite{Williamson2020,Jalas2013}. Bulk optical isolators are typically based on magneto-optical effect \cite{Aplet,Levy2002} in gallium or iron garnets. Analogous magneto-optical isolators have also been developed for integrated photonics by integrating yttrium iron garnet (YIG) thin films on various photonics platforms \cite{Auracher1975,Shintaku1998,Fujita2000,Levy2002,Shoji2007,Ross2014,Ross2019}. However, these isolators exhibit high losses due to challenges in the on-chip integration of high-quality YIG thin films. In addition, the isolation bandwidth is limited due to the use of interferometric structures to convert non-reciprocal phase shifts into amplitude.

Besides the magneto-optical effect, nonlinear optical processes have been used to demonstrate on-chip optical isolation \cite{Fan2012,Li2020c,White2022,Kim2015,Dong2015c,Sohn2019,Hafezi2012,Shen2016a,Fang2017a}. Moreover, traveling-wave modulation can also be used for on-chip isolators by introducing non-reciprocal mode conversion and amplitude modulation \cite{Williamson2020,Yu2009, Herrmann2021,Yu2022}. These approaches all require stringent phase-matching, offering limited bandwidths. In contrast, standing-wave modulation in tandem configuration can exhibit large isolation bandwidth, which is determined by the modulator's optical wavelength range \cite{Doerr2014}. In particular, this method is well-suited for thin-film lithium niobate (LN) photonics due to its strong intrinsic electro-optic response~\cite{Zhu2021}.

\begin{figure*}[htb]
    \centering
    \includegraphics[width=0.9\textwidth]{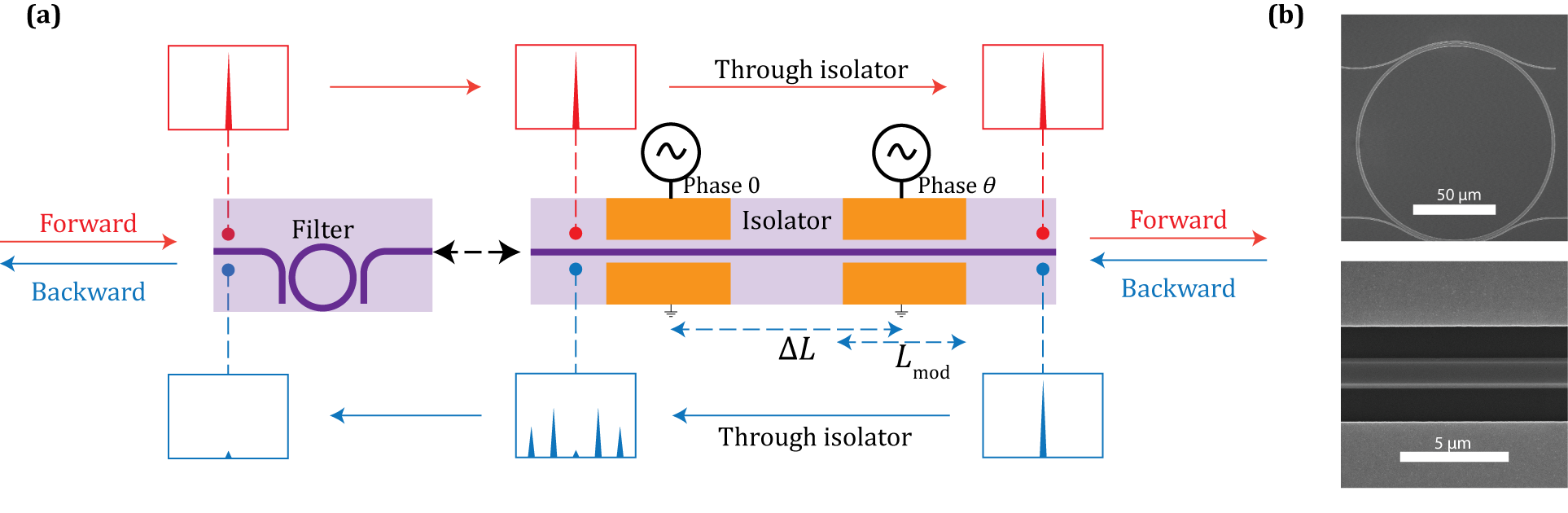}
    \caption{(a) Schematic of the optical isolator. The spectra in red (blue) show the transformation of the optical signal through the isolator in the forward (backward) propagation direction. An add-drop micro-ring resonator filter device is placed at the input of the isolator for sideband suppression in the backward direction. (b) Scanning electron microscopy images showing the fabricated micro-ring filter and electro-optic modulator on thin film LN.}
    \label{fig1}
\end{figure*}

In this letter, we demonstrate the broadband optical isolation based on the standing-wave electro-optic modulation with tandem configuration \cite{Doerr2011,Lin2019} on thin-film LN integrated photonics. The schematic of the optical isolator is shown in Fig.~\ref{fig1}a. Two standing-wave phase modulators are separated by an optical delay line ($\Delta T$). The modulators are driven at microwave frequency $F_\mathrm{m}$ with a relative microwave phase shift $\theta$. In the forward direction, the time-dependent transmission is
\begin{equation}
    T_{\mathrm f}(t) = e^{i2A \,\cos{\Big[ 2\pi F_{\rm m} t + \Big(\frac{\theta - 2\pi F_m \Delta T}{2}\Big)\Big]} \cos{\Big(\frac{\theta + 2\pi F_{\rm m} \Delta T}{2}\Big)}}
\end{equation}
where $A$ is the modulation amplitude. In the backward direction, the transmission becomes
\begin{equation}
    T_{\mathrm b}(t) = e^{i2A \,\cos{\Big[ 2\pi F_{\rm m} t + \Big(\frac{\theta - 2\pi F_{\rm m} \Delta T}{2}\Big)\Big]} \cos{\Big(\frac{\theta - 2\pi F_{\rm m} \Delta T}{2}\Big)}}.
\end{equation}
If the microwave phase shift $\theta$ and the delay line $\Delta T$ are adjusted such that ($\theta+2\pi F_{\rm m}\Delta$T$)=\pi$, we always have $T_{\mathrm f}(t)=1$ and the forward optical field is transmitted without attenuation. In contrast, the backward optical field is phase modulated -
\begin{equation}
    T_{\mathrm b}(t) = \sum_k i^k \,J_k \Bigg[2A\,\cos{\Big(\frac{\theta - 2\pi F_{\rm m} \Delta T}{2}\Big)}\Bigg] \, e^{ik\big(2\pi F_{\rm m} t + \frac{\theta - 2\pi F_{\rm m} \Delta T}{2}\big)}
\end{equation}
where the amplitude for each modulation sideband is given by the $k^{\mathrm{th}}$-order Bessel function of the first kind ($J_k$). If the power at the center frequency vanishes
\begin{equation}
\label{eq4}
    J_0 \Bigg[2A\,\cos{\Big(\frac{\theta - 2\pi F_{\rm m} \Delta T}{2}\Big)}\Bigg] = 0,
\end{equation}
a bandpass filter centered at the input optical frequency completely prevents the backward propagation (Fig.~\ref{fig1}a). The microwave drive and optical delay line are typically adjusted such that is $\theta = 2\pi F_{\rm m}\Delta T = \pi/2$. With modulation amplitude $A = 0.385\pi$, the backward propagation is completely suppressed ($J_0 [2A] = 0$). The corresponding optical delay length is $\Delta L = c/4 F_\mathrm{m} n_\mathrm{g}$, where $c$ is the vacuum speed of light, and $n_\mathrm{g}$ is the optical group refractive index.

\begin{figure}[htb]
    \centering
    \includegraphics[width=0.45\textwidth]{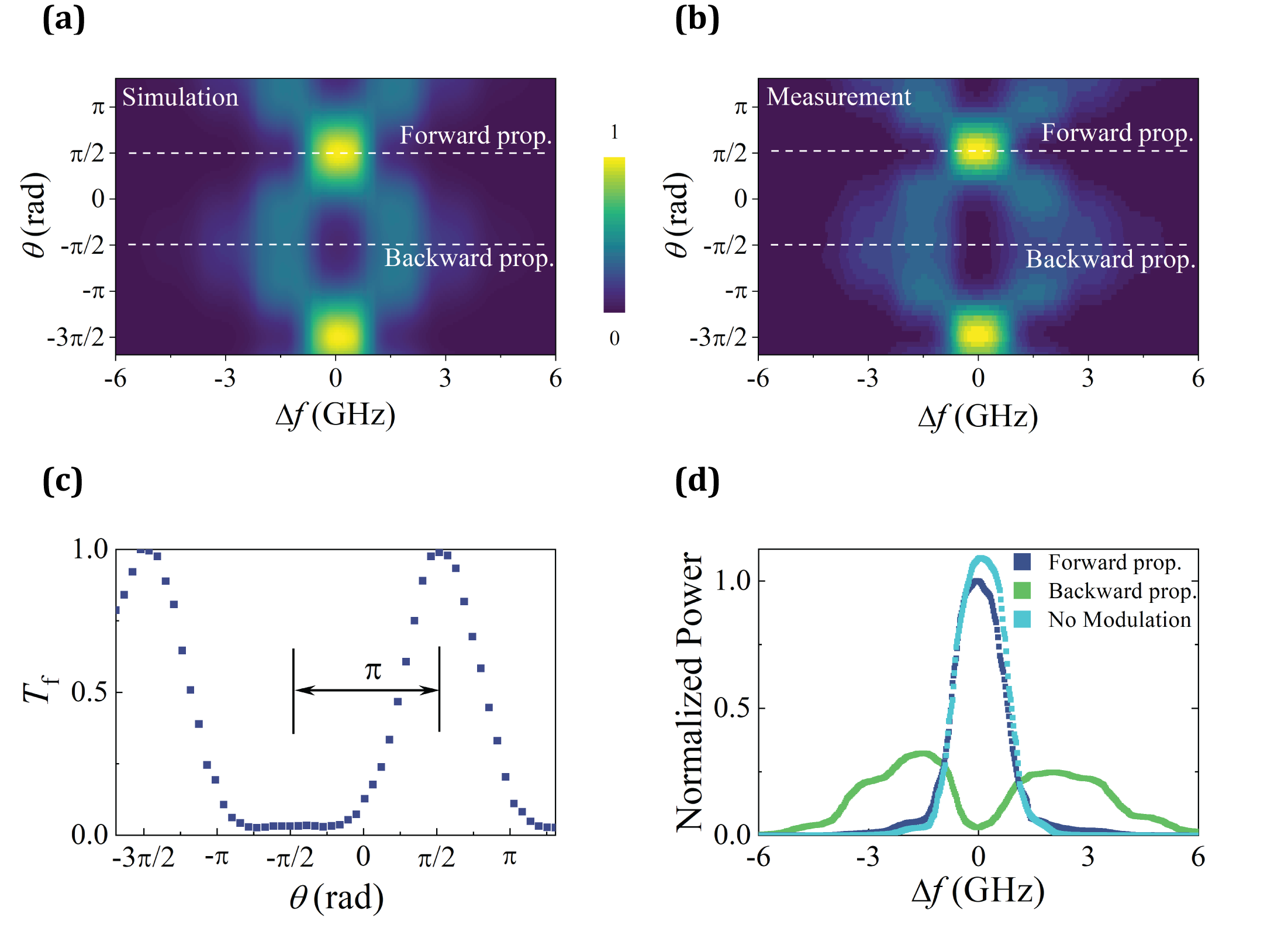}
    \caption{(a) Calculated optical spectrum with varied $\theta$. Optical frequency ($\Delta f$) is offset by the input light frequency. (b) Measured optical spectrum with OSA for varied $\theta$. (c) Measured center frequency transmission for forward propagation against varied $\theta$. (d) Measured optical spectrum in the forward and backward directions showing 15~dB isolation. Reference transmission for forward propagation without modulation is included for comparison.}
    \label{fig2}
\end{figure}

The isolator is fabricated on 600 nm X-cut thin-film LN-on-insulator substrate (Fig.~\ref{fig1}b). The fabrication procedure is similar to our previous works~\cite{briggs2021simultaneous, xu2021bidirectional, chen2022ultra}. The waveguide pattern is first written with 100-kV electron-beam lithography (EBL) on hydrogen silsesquioxane (HSQ) resist. After developing the exposed pattern in Tetramethylammonium Hydroxide (TMAH), Argon-based plasma etching is used to transfer the waveguide pattern from HSQ to the thin-film LN device layer. The redeposited LN from the plasma etching is removed by submerging the chip in a mixture of hydrogen peroxide and ammonium hydroxide. Finally, the chip is annealed to reduce propagation loss. The electrode pattern is written using EBL in polymethyl methacrylate (PMMA) resist. After development in a mixture of Methyl isobutyl ketone (MIBK) and isopropyl alcohol (IPA), a 100-nm layer of Au is deposited over the developed pattern, followed by metal liftoff in acetone.

To use the largest electro-optic coefficient ($r_{33}$) for modulation, the waveguide is aligned along the Y direction of thin-film LN, and the fundamental transverse-electric (TE) optical mode is used. The length of each modulator is $L_\mathrm{mod}=1$~cm, and the gap between the signal and ground electrodes is 5~$\mu$m. We determine a half-wave voltage (V$_{\pi}$) of 6.5 V and a microwave bandwidth of 2 GHz for the modulators. The optical waveguide width is 1~$\mu$m, resulting in a group refractive index of $n_\mathrm{g}=2.24$ for the fundamental TE mode at wavelength $\lambda=1550$~nm. To match the $\pi/2$ microwave phase shift, the length of the optical delay line between the modulators (center-to-center) is fixed at $\Delta L = 2.23$~cm. A variable microwave phase shifter is introduced between the two modulator drivers to precisely tune $\theta$.

We first characterize the dependence of the tandem modulator on the microwave phase shift $\theta$. The thin-film LN waveguide is cleaved and coupled to a pair of lens fibers. A continuous-wave laser at 1550 nm is launched into the device, and the output is measured with an optical spectrum analyzer (OSA). The modulators are driven by an amplified microwave signal at 1.5 GHz, split into two paths with a 3 dB power splitter. The driving voltage is fixed at 5 V, corresponding to $A\approx0.385\pi$ at 1550 nm. As shown in  Fig.~\ref{fig2}b, optical power oscillates between the center frequency and sidebands on changing the phase shift $\theta$. The phase change period is $\pi$, which agrees with the theoretical calculation using Eq.~3 (Fig.~\ref{fig2}a). The broadening of center frequency at phase shift $\pi/2$ in Fig.~\ref{fig1}b is due to the finite OSA resolution ($\sim$1.25 GHz at 1550 nm wavelength). Therefore, we include the influence of the finite OSA resolution in the calculated result in Fig.~\ref{fig2}a for the direct comparison with measured data.

The measured power at the center frequency with varied phase shift $\theta$ is shown in Fig.~\ref{fig2}c. To achieve the optimum isolation performance, the forward and backward propagation optical fields should experience $\pi/2$ and $-\pi/2$ microwave phase delay, respectively. We fixed the microwave phase shift at $\pi/2$ and measured the optical transmission spectrum along forward and backward directions (Fig.~\ref{fig2}d). The optical power at the center frequency in the backward direction is more than 15 dB lower than in the forward direction. We also measure the transmission spectrum along the forward direction without modulation as a reference. A slight decrease of the optical power along the forward direction is observed compared with the optical power measured without modulation, indicating an insertion loss below 0.5 dB for the isolator. 

\begin{figure}[htb]
    \centering
    \includegraphics[width=0.45\textwidth]{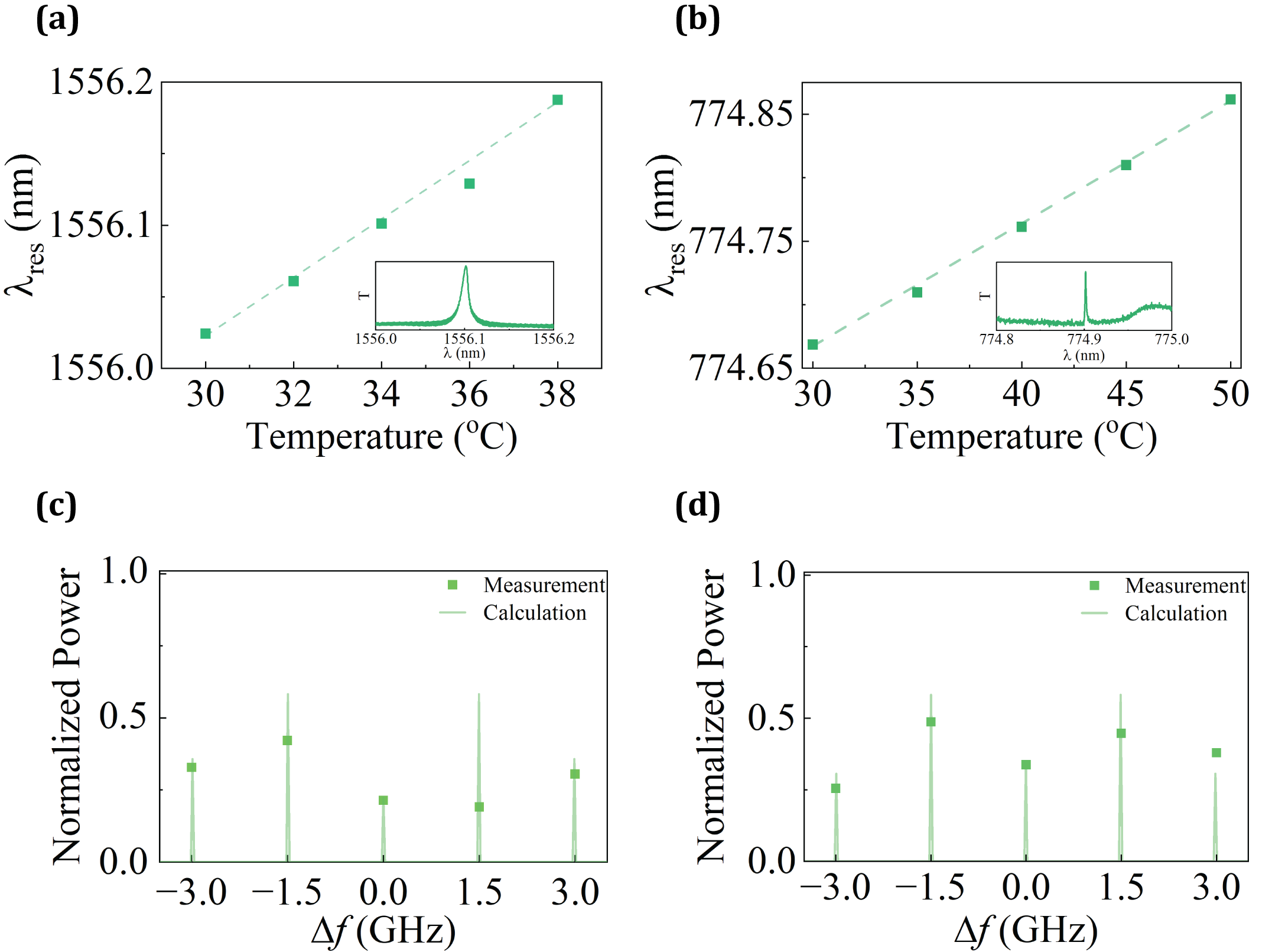}
    \caption{(a) Resonance tuning of micro-ring resonator in telecom regime. Inset: transmission spectrum at room temperature. (b) Resonance tuning of micro-ring resonator in visible regime. Inset: transmission spectrum at room temperature. (c) Normalized telecom and (d) visible transmission for backward propagation captured by tuning the micro-ring filter passband.}
    \label{fig3}
\end{figure}

\begin{figure}[htb]
    \centering
\includegraphics[width=0.45\textwidth]{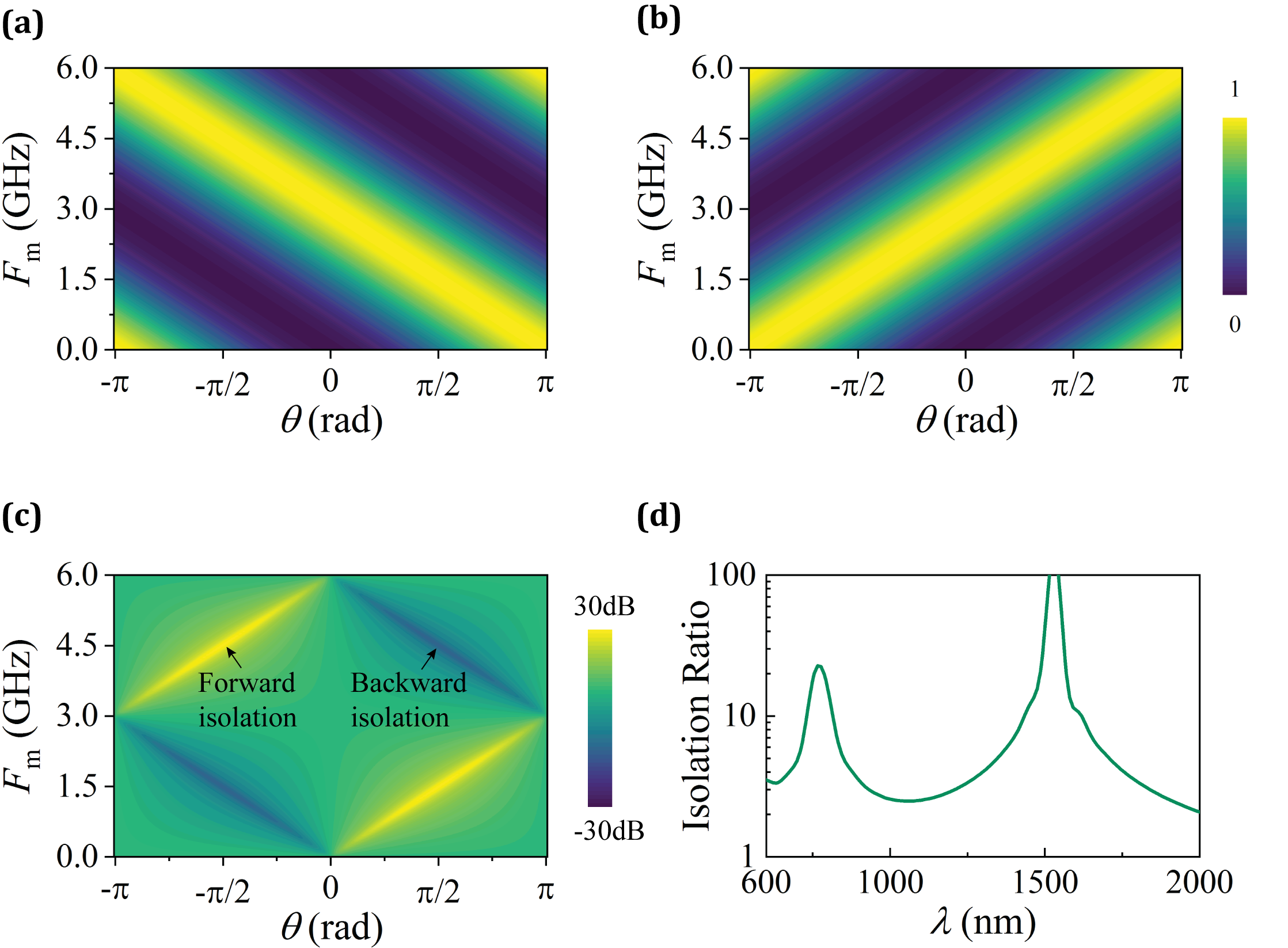}
    \caption{(a) Forward transmission, (b) Backward transmission, and (c) Isolation ratio for varied $F_{\mathrm{m}}$ and $\theta$. (d) Calculated isolation ratio for specified device configuration at different wavelengths.}
    \label{fig4}
\end{figure}

To realize the complete isolator device on a chip, we further integrate micro-ring resonators on thin-film LN (Fig.~\ref{fig1}b). The narrow linewidth of micro-ring resonators makes them good band-pass filters in the add-drop configuration. The micro-ring waveguide width is 1.5~$\mu$m with a radius of 60~$\mu$m, and the bus waveguides are 700~nm wide. The measured resonance linewidth is $\sim$1.5 GHz near 1550~nm. We measured the backward propagation transmission spectrum at different filter center wavelengths (Fig.~\ref{fig3}c). We tune the center wavelength of the micro-ring bandpass filter by changing the device temperature (Fig.~\ref{fig3}a). When the center wavelength is aligned with the wavelength of the input optical field, an isolation ratio of 7 dB is observed. Compared with the result measured by OSA (Fig.~\ref{fig2}d), the isolation ratio is degraded due to the larger bandwidth of the bandpass filter.

Given LN's large transparency window and the broad optical bandwidth of standing wave modulators, this isolator may be used across a broad wavelength range. To demonstrate the large optical bandwidth, we test the same isolator at $\lambda=775$~nm. The modulation frequency, microwave phase shift, and amplitude are kept unchanged. The micro-ring exhibits a resonance linewidth of $\sim$1 GHz near 775 nm. Analogous to $\lambda=1550$~nm, the center wavelength can be tuned with temperature (Fig.~\ref{fig3}b). When the center wavelength is aligned with the wavelength of the input optical field, an isolation ratio of 5 dB is observed near 775 nm (Fig.~\ref{fig3}d).

While the bandpass filter bandwidth is smaller at $\lambda=$775~nm wavelength, the isolation ratio is still lower than that for 1550~nm. This is due to the change in the group refractive index $n_\mathrm{g}$, which leads to the mismatch between the microwave phase shift $\theta$ and optical delay $\Delta T$. While the optical delay line cannot be tuned, optimal isolation can be easily achieved by adjusting the microwave drive. The optical transmission along the forward direction is always unity provided $\theta+2\pi F_\mathrm{m}\Delta T = \pi$ (Fig.~\ref{fig4}a). By choosing $\theta$ and $F_\mathrm{m}$ such that $\theta-2\pi F_\mathrm{m}\Delta T = \pi$ (Fig.~\ref{fig4}b), the power in the center frequency along the backward propagation can be suppressed. The calculated isolation ratio is shown in Fig.~\ref{fig4}c. Besides isolation along the forward direction, we can also invert the isolation direction by changing the microwave phase shift $\theta$ into $-\theta$. In addition, optical isolation can still be achieved across a large wavelength range without adjusting microwave drive signals. While this configuration is optimized to achieve a high isolation ratio and low insertion loss at $\lambda=1550$~nm, the isolation ratio remains above 10 dB in an optical bandwidth over 100~nm (Fig.~\ref{fig4}d). Moreover, the isolator performs well at 775~nm, with a calculated isolation ratio above 12 dB and a 10 dB optical bandwidth of over 80~nm. Such a dual-band isolator is ideal for nonlinear optical applications such as second harmonic generation and parametric down-conversion.

In conclusion, using thin-film LN photonics, we have demonstrated integrated optical isolators based on standing-wave modulation in the tandem configuration. As no stringent phase-matching condition is required, the isolator bandwidth is only restricted by the wavelength range of modulators, which can span across the visible and near-infrared spectrum. We verified the large operational wavelength range by demonstrating optical isolation in telecom and visible regimes. We experimentally measure maximum isolation over 15 dB at $\lambda=1550$~nm with insertion loss smaller than 0.5 dB. The condition to achieve optimum isolation at different wavelengths is discussed. Microwave drive frequency and phase shift should be adjusted to compensate for the optical group velocity dispersion. The isolation performance can be further improved using bandpass filters with a higher extinction ratio and narrower bandwidth. The flexibility and tunability offered by this isolator can be utilized to demonstrate ultra-broadband isolation on integrated photonic circuits.

\begin{backmatter}
\bmsection{Funding} Authors acknowledge the support from U.S. Department of Energy, Office of Advanced Scientific Computing Research (Field Work Proposal ERKJ355); Office of Naval Research (N00014-19-1-2190); National Science Foundation (CCF-1907918, ITE-2134830); Coherent/II-VI foundation.

\bmsection{Acknowledgments} M.S. acknowledges helpful discussions with Dr. Liang Zhang, Chaohan Cui, Dr. Roland Himmelhuber, and Dr. Abhinav Nishant. Device fabrication is performed in the OSC cleanrooms at the University of Arizona.
\bmsection{Disclosures} The authors declare no conflicts of interest.
\bmsection{Data availability} Data underlying the results presented in this paper is not publicly available at this time but may be obtained from the authors upon reasonable request.

\end{backmatter}

\bibliography{NonReciprocity.bib}

\bibliographyfullrefs{NonReciprocity.bib}

\end{document}